# 基于改进最大相关最小冗余判据的暂态稳定评估特征选择

李扬，顾雪平

(华北电力大学电气与电子工程学院，河北省 保定市 071003)

## Feature Selection for Transient Stability Assessment Based on Improved Maximal Relevance and Minimal Redundancy Criterion

LI Yang, GU Xueping

(School of Electrical and Electronic Engineering, North China Electric Power University, Baoding 071003, Hebei Province, China)

**ABSTRACT:** A new feature selection method based on an improved maximal relevance and minimal redundancy (mRMR) criterion was proposed for power system transient stability assessment. First, the standard mRMR was improved by introducing a weight coefficient in the evaluation criteria to refine the measurement of the features correlation and redundancy. Then, the possible real-time information provided by phasor measurement unit (PMU) considered, a group of system-level classification features were extracted from the power system operation parameters to build the original feature set, and the improved mRMR was employed to evaluate the classification capability of the original features for feature selection. A group of nested candidate feature subsets were obtained by using the incremental search technique, and each candidate feature subset was tested by a support vector machine classifier to find the optimal feature subset with the highest classification accuracy. The effectiveness of the proposed method was validated by the simulation results on the New England 39-bus system and IEEE 50-generator test system.

**KEY WORDS:** transient stability assessment; feature selection; maximal relevance and minimal redundancy (mRMR); support vector machine (SVM); phasor measurement unit (PMU)

**摘要**：提出一种基于改进最大相关最小冗余判据(maximal relevance and minimal redundancy，mRMR)的暂态稳定评估特征选择方法。首先对标准 mRMR 方法进行改进，在最大相关、最小冗余判据中引入了一个权重因子以细化对特征相关性和冗余性的度量。然后，考虑相量测量单元可以提供的故障后实测信息，构造由系统特征构成的原始特征集，将改进的 mRMR 应用于特征选择。通过增量搜索算法得到一组嵌套的候选特征子集，并使用支持向量机分类器验证各候选特征子集的分类性能，选择得到具有最大分类正确率的特征子集。基于新英格兰 39 节点系统和 IEEE 50 机测试系统的算例结果验证了所提特征选择方法的有效性。

**关键词**：暂态稳定评估；特征选择；最大相关最小冗余；支持向量机；相量测量单元

## 0 引言

电力系统暂态稳定评估(transient stability assessment，TSA)是保证电力系统安全稳定运行的基础性手段[1]。近年来随着计算技术的快速发展，基于人工神经网络(artificial neural network，ANN)[2-4]、和支持向量机(support vector machine，SVM)[5-6]等模式识别技术的 TSA(pattern recognition-based TSA，PRTSA)受到了各国学者的广泛关注，取得了良好的效果。PRTSA 无需建立系统的数学模型，它的主要任务是建立系统变量和系统稳定结果间的关系映射[3]，包括特征抽取与选择、训练和测试样本集的建造、分类器的设计和训练、分类性能评估等，其中输入特征选择和输入空间降维是 PRTSA 的首要问题[2]。由模式识别理论可知，一个模式分类器的输入特征过多会增加知识发现的计算成本、降低训练模型的精度甚至导致"维数灾难"问题[7]，同时已经证明最优特征子集选择是 NP 难问题[8]，而电力系统的高维性又是理论研究和工程实践中重要难题[9]，因此，特征选择是 PRTSA 研究和应用所亟待解决的问题。

针对暂态稳定评估的特征选择问题，国内外学者已进行了一些有益的探索[2,10-12]。文献[10]采用 Fisher 线性判别函数和序列特征选择技术进行神经网络训练特征的选择，将其应用于 IEEE 50 机测试

系统的仿真结果验证了所提方法的有效性；文献[2]提出基于粗糙集理论特征离散化的类别可分离性判据，并利用 Tabu 搜索技术从初始特征集中选择出一组有效特征，显著减低了输入空间维数；文献[11]提出了一种基于输入空间的可分性评估的特征选择方法，并采用广度优先搜索寻找最优的特征子集；文献[12]提出基于支持向量机的双阶段特征选择方法，先以支持向量机递归特征选择法对原始特征集降维，再以支持向量机为分类器的包装法，用最佳优先搜索算法得到一组近似最优特征子集。同时，近年来相量测量单元(phasor measurement unit，PMU)的引入使获取同步的故障后实时信息成为现实，为暂态稳定分析带来新的思路和契机[13-14]。

文献[15]提出了一种用于模式识别特征选择的最大相关最小冗余准则(maximal relevance and minimal redundancy，mRMR)，其核心思想是从给定的特征集合中寻找与目标类别有最大相关性且相互之间具有最少冗余性的特征子集。本文对 mRMR 进行改进，并将其应用于暂态稳定评估特征选择。考虑 PMU 可以提供的故障后实测信息，在构造由系统特征构成的原始特征集的基础上，提出一种新的暂态稳定评估特征选择方法——基于改进 mRMR 和支持向量机的特征选择法。该方法引入特征相关冗余权重因子对标准 mRMR 进行改进，并以改进 mRMR 为评价判据、通过增量搜索算法得到一组嵌套的候选特征子集；然后使用支持向量机分类器验证各候选特征子集的优劣，选择得到最大分类正确率的特征集合。

# 1 mRMR 简介

## 1.1 最大相关最小冗余准则

给定两个随机变量 $x$ 和 $y$，其概率密度为 $p(x)$ 和 $p(y)$，联合概率密度为 $p(x,y)$，则 $x$ 和 $y$ 之间的互信息定义为

$$I(x;y) = \iint p(x,y)\log\frac{p(x,y)}{p(x)p(y)}\mathrm{d}x\mathrm{d}y \quad (1)$$

最大相关和最小冗余的测度指标分别定义为

$$\max D(S,c), D = \frac{1}{|S|}\sum_{x_i \in S} I(x_i;c) \quad (2)$$

$$\min R(S), R = \frac{1}{|S|^2}\sum_{x_i,x_j \in S} I(x_i;x_j) \quad (3)$$

式中：$S$ 和 $|S|$ 分别为特征集合及其包含的特征数目；$c$ 为目标类别；$I(x_i;c)$ 为特征 $i$ 和目标类别 $c$ 之间的互信息；$I(x_i;x_j)$ 为特征 $i$ 与特征 $j$ 之间的互信息；$D$ 特征集 $S$ 中各特征 $x_i$ 与类别 $c$ 之间互信息的均值，表示特征集与相应类别的相关性；$R$ 为 $S$ 中特征间互信息的大小，表示特征之间的冗余性。

特征选择的目标是期望所选特征子集的分类性能最高、同时特征维数尽量少，这就要求特征集与类别间相关性最大、特征之间冗余性最小。综合考虑上述两个测度指标，得到最大相关最小冗余准则如下：

$$\max \Phi(D,R), \Phi = D - R \quad (4)$$

## 1.2 增量搜索算法

在实际应用中，可以采用增量搜索算法来选取由 $\Phi(\cdot)$ 所定义的近似最优特征。假设原始特征集为 $X$，已选择包含 $m-1$ 个特征的特征子集为 $S_{m-1}$，则特征选择的任务就是从剩余的特征集 $\{X-S_{m-1}\}$ 中选择第 $m$ 个特征使得式(4)中的 $\Phi(\cdot)$ 最大化，这个特征应满足：

$$\max_{x_j \in X-S_{m-1}}[I(x_j;c) - \frac{1}{m-1}\sum_{x_i \in S_{m-1}} I(x_j;x_i)] \quad (5)$$

# 2 基于改进 mRMR 和 SVM 的特征选择方法

## 2.1 对 mRMR 的改进

针对标准 mRMR 准则直接将最大相关和最小冗余作差、难以细致刻画特征相关性和冗余性权重的不足，本文引入特征相关冗余权重因子 $\alpha$ 以细化对特征相关性和冗余性的度量。通过对 $\alpha$ 进行不同的赋值，对准则中特征相关性与冗余性的加权系数进行调整，以求得到最好的特征选择结果，相应地将式(4)的最大相关最小冗余准则修正为

$$\max \Phi(D,R), \Phi = \alpha D - (1-\alpha)R \quad (6)$$

式中当 $\alpha=0.5$ 时，式(6)即退化为式(4)的标准 mRMR 准则。

## 2.2 基于改进 mRMR 的特征初选

### 2.2.1 特征选择流程

基于 mRMR 和 SVM 的特征选择方法共分为 2 步：1) 引入特征相关冗余权重因子对标准 mRMR 进行改进，并使用改进 mRMR 法生成候选特征子集，避免了对原始特征集进行穷尽式搜索；2) 使用支持向量机分类器验证各候选特征子集的优劣，选择得到最大分类正确率的特征子集。所提方法的详细流程如图 1 所示，图中 $N$ 为原始特征集的特征

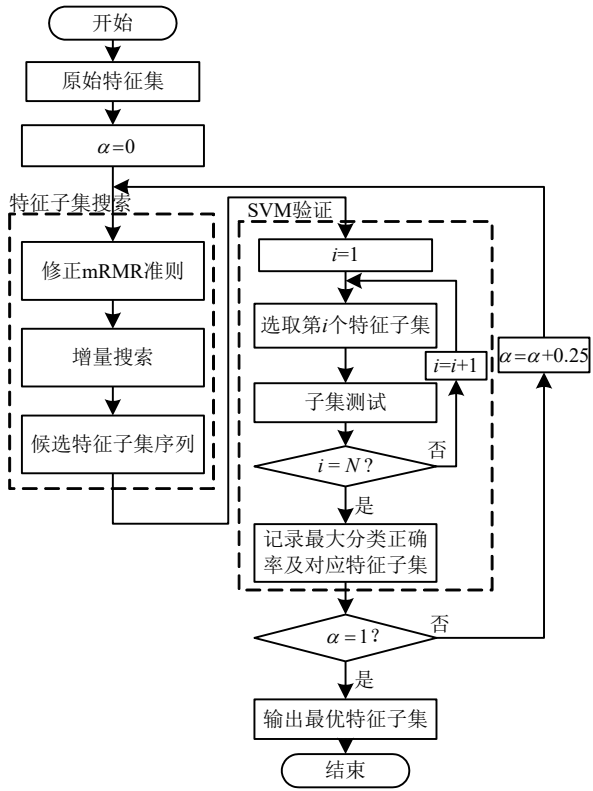

**图 1 基于改进 mRMR 和 SVM 的特征选择流程**
**Fig. 1 Flowchart of feature selection based on improved mRMR and SVM**

数量，α为特征相关冗余权重因子。

#### 2.2.2 生成候选特征子集

设 $X$ 为共有 $N$ 个特征的原始特征集，$S$ 为已选特征集，$F$ 为待选特征集，则具体步骤如下：

1）初始化。$X \to F$，$\varnothing \to S$。

2）计算 $\forall x_i \in F$ 与目标类 $c$ 间的相关性 $I(x_i; c)$，找到使 $\max[I(x_i; c)]$ 成立的特征并记为 $x^*$，令 $F-\{x^*\} \to F_1$，$\{x^*\} \to S_1$。

3）设 $x_i \in F_{m-1}$，$x_j \in S_{m-1}(m=2,\ldots,N)$，从 $F_{m-1}$ 中找到使式(6)成立的最大特征并记为 $x^{**}$，令 $F_{m-1}-\{x^{**}\} \to F_m$，$S_{m-1}+\{x^{**}\} \to S_m$。

4）重复步骤3），直到 $F$ 为 $\varnothing$，则得到的备选特征集 $S$，并将 $S$ 中的特征按照 $\Phi(D,R)$ 降序排列，得到 $N$ 个嵌套的候选特征子集 $S_1 \subset S_2 \subset \ldots \subset S_N$。

### 2.3 SVM 验证

SVM 是在统计学习理论基础上发展起来的一种模式识别方法[16]，在解决小样本、非线性和高维数的分类问题上具有良好的表现。它通过核函数进行空间映射并构造最优分类超平面解决分类器的构造问题。

本文采用 SVM 分类器验证各特征子集的性能，其模型参数包括正则化参数和核参数。文中核函数选用径向基核函数，即

$$K(\boldsymbol{X}_i, \boldsymbol{X}_j) = \exp(-\gamma \|\boldsymbol{X}_i - \boldsymbol{X}_j\|^2), \gamma > 0 \quad (7)$$

式中：$\boldsymbol{X}_i$ 和 $\boldsymbol{X}_j$ 分别表示第 $i$ 和 $j$ 个样本的输入特征向量；$\gamma$ 为核参数。

分类正确率 $J(S)$ 是评估特征子集 $S$ 优劣的重要指标。由于不知道 SVM 在各特征子集上的实际正确率，所以 $J(S)$ 取为特征集 $S$ 的训练集上 5-折交叉验证 SVM 评估正确率的平均值，其无偏估计性抑制了训练过程中"过拟合"和"欠拟合"。设 $F^*$ 为最大分类正确率的特征集合，则此部分算法描述如下：

1）将权重因子 $\alpha$ 在[0,1]区间内、以 0.25 为步长赋值，即 $\alpha_i = 0, 0.25, \ldots$，$1 \leq i \leq 5$，采用改进 mRMR 分别得到与权值 $\alpha_i$ 对应的一组嵌套的候选特征集 $S_i^1, S_i^2, \ldots \ldots$，$1 \leq j \leq N$。

2）对每个权值 $\alpha_i$ 所对应的嵌套候选特征集组，使用 SVM 分类器以每次递增 1 个特征的方式计算该组各特征子集 $S_i^j$ 的分类正确率 $J(S_i^j)$，并记录其中的最大分类正确率 $e_i^* = \max\{J(S_i^j)\}$ 及其对应的特征子集。

3）比较所有不同权值下的子集测试结果，令 $F^* = A_{\max}\{e_i^*\}$，其中 $A_{\max}\{\cdot\}$ 表示分类正确率 $e_i^*$ 最大时所对应的特征子集。如果有多个特征集合同时取得最大分类正确率，则 $F^*$ 取其中特征数量最少的特征集合。至此，$F^*$ 即为所求的最优特征子集。

## 3 原始特征集的构造

TSA 的原始特征可以从以下角度进行分类：从是否随系统规模变化上分，原始特征可分为单机特征和系统特征[2]；从时间上分，原始特征可分为静态特征和动态特征[12]；从空间上分，原始特征可分为电网参数特征和发电机参数特征[12]。近年来，PMU 的引入使获取同步的故障后实时信息成为现实，从而为 PRTSA 提供了新的输入特征。

本文在原始特征集的构建过程中，依据了以下原则：1）主流性原则，通过对电力系统暂态稳定物理过程的本质特性做深入分析，求取与稳定性强相关的输入特征量；2）实时原则，基于引入 PMU 后的优势，求取来自扰动发生之后的信息，可以动态实时的表征故障发生后系统的运行状态；3）系统性原则，选用系统特征，而不是单机特征，以保证输入变量个数不随系统规模的增大成比例增长，从而适合大系统的稳定分析。

本文按照上述原则，在综合现有的研究文献的

基础上，考虑 PMU 可能提供的故障后实测信息，通过大量仿真分析，构建了一组由 33 个系统特征所组成的原始特征集，如表 1 所示。表中 $t_0$ 为故障初始时刻，$t_{cl}$ 为故障切除时刻，$t_{cl+3c}$ 为故障切除后第 3 周波，$t_{cl+6c}$ 为故障切除后第 6 周波，$t_{cl+9c}$ 为故障切除后第 9 周波。

表 1 中，Tz1 反映了系统的整体静态稳定水平；Tz2—Tz4 为由故障瞬间抽取的特征量，其中 Tz2 反映了受扰最严重的发电机的失稳趋势，Tz3 反映了受扰最严重发电机的静态运行点，Tz4 反映了系统中各发电机故障瞬间供求关系失衡的平均水平；Tz5—Tz12 为由故障切除时刻抽取的特征量，其中 Tz5 反映了该时刻扰动对系统的破坏情况，Tz7 反映了该时刻转角最领先发电机的失稳趋势，Tz8 反映了具有最大动能发电机在故障切除后的减速能力；Tz13—Tz33 为考虑 PMU 可以提供故障后实测信息的优势而提出的特征量，它们反映了故障切除后的过程中系统的稳定特性。因此，上述特征量比较全面地表征了系统在故障前后的整个受扰过程中不同侧面、不同阶段的稳定特性、而彼此间又相互补充，故而选其构成系统暂态受扰模式空间的原始特征集。

## 4 算例分析

### 4.1 新英格兰 39 节点系统

#### 4.1.1 算例介绍

新英格兰 39 节点系统共由 10 台发电机、39 母线和 46 条线路所组成[2,12]，其中 39 号母线所连的发电机为外网等值机，其单线结构如下图 2 所示。

表 1 数据集的输入特征量
Tab. 1 Input features of data set

| 编号 | 输入特征量 |
| --- | --- |
| Tz1 | 系统中各发电机机械功率的平均值 |
| Tz2 | $t_0$ 时刻所有发电机初始加速度的最大值 |
| Tz3 | $t_0$ 时刻具有最大加速度发电机的初始角度 |
| Tz4 | $t_0$ 时刻所有发电机初始加速功率的均值 |
| Tz5 | $t_{cl}$ 时刻系统冲击的大小 |
| Tz6 | $t_{cl}$ 时刻与惯性中心相差最大的发电机转子角度 |
| Tz7 | $t_{cl}$ 时刻具有最大转角发电机的动能 |
| Tz8 | $t_{cl}$ 时刻具有最大动能发电机的转子角度 |
| Tz9 | $t_{cl}$ 时刻所有发电机转子动能的最大值 |
| Tz10 | $t_{cl}$ 时刻所有发电机转子动能的平均值 |
| Tz11 | $t_{cl}$ 时刻发电机转子最大相对摇摆角 |
| Tz12 | $t_{cl}$ 时刻与惯性中心相差最大的发电机角速度 |
| Tz13 | $t_{cl+3c}$ 时刻系统冲击的大小 |
| Tz14 | $t_{cl+3c}$ 时刻所有发电机转子动能的最大值 |
| Tz15 | $t_{cl+3c}$ 时刻所有发电机转子动能的平均值 |
| Tz16 | $t_{cl+3c}$ 时刻与惯性中心相差最大的发电机转子角度 |
| Tz17 | $t_{cl+3c}$ 时刻发电机转子最大相对摇摆角 |
| Tz18 | $t_{cl+3c}$ 时刻具有最大转角发电机的动能 |
| Tz19 | $t_{cl+3c}$ 时刻与惯性中心相差最大的发电机角速度 |
| Tz20 | $t_{cl+6c}$ 时刻系统冲击的大小 |
| Tz21 | $t_{cl+6c}$ 时刻所有发电机转子动能的最大值 |
| Tz22 | $t_{cl+6c}$ 时刻所有发电机转子动能的平均值 |
| Tz23 | $t_{cl+6c}$ 时刻具有最大转角发电机的动能 |
| Tz24 | $t_{cl+6c}$ 时刻与惯性中心相差最大的发电机转子角度 |
| Tz25 | $t_{cl+6c}$ 时刻发电机转子最大相对摇摆角 |
| Tz26 | $t_{cl+6c}$ 时刻与惯性中心相差最大的发电机角速度 |
| Tz27 | $t_{cl+9c}$ 时刻系统冲击的大小 |
| Tz28 | $t_{cl+9c}$ 时刻具有最大转角发电机的动能 |
| Tz29 | $t_{cl+9c}$ 时刻所有发电机转子动能的最大值 |
| Tz30 | $t_{cl+9c}$ 时刻所有发电机转子动能的平均值 |
| Tz31 | $t_{cl+9c}$ 时刻与惯性中心相差最大的发电机转子角度 |
| Tz32 | $t_{cl+9c}$ 时刻发电机转子最大相对摇摆角 |
| Tz33 | $t_{cl+9c}$ 时刻与惯性中心相差最大的发电机角速度 |

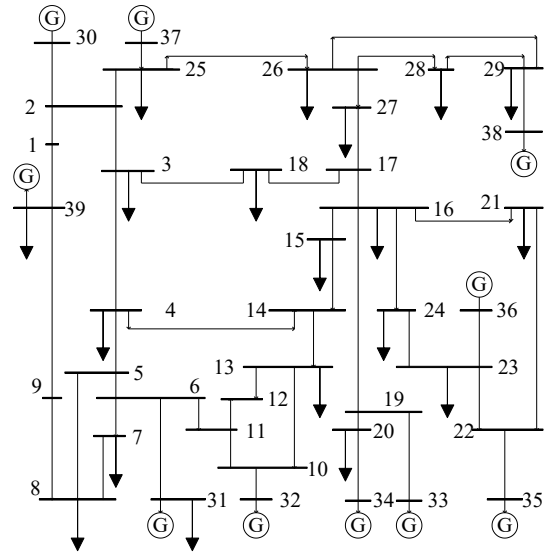

图 2 新英格兰 39 节点系统
Fig. 2 New England 39-bus system

#### 4.1.2 样本集的构造

发电机模型采用 4 阶模型，各发电机(除等值机外)的励磁均采用 IEEE DC1 型励磁系统模型，负荷模型为恒阻抗模型；故障类型为三相短路，故障清除时间为 0.1 s，故障清除后系统拓扑结构不变；系统在从 80%以 10%递增到 130%共计 6 个负荷水平下，每个负荷水平下随机设置 5 种发电机出力，选择 20 个不同的故障位置；按在仿真结束时，任意两台发电机的最大相对功角差是否大于 360°来判定系统是否失稳[6,12]；仿真软件为 PSD 电力系统分析软件(中国版 BPA)，共生成 600 个样本，随机选取其中 400 个构成训练集，其余为测试集。

### 4.1.3 结果与讨论

基于所构建的原始特征集，采用改进 mRMR 求取不同权重因子 $\alpha$ 对应的序列特征子集并按 $\Phi(D,R)$ 降序排列，结果如表 2 所示。

**表 2 特征降序排列分布**
**Tab. 2 Feature distribution in descending order**

| 权重因子 | 特征排序 |
| --- | --- |
| 0 | 32, 10, 4, 19, 1, 9, 28, 16, 18, 33, 23, 27, 3, 21, 15, 11, 12, 31, 17, 8, 6, 13, 5, 29, 14, 7, 22, 26, 30, 25, 2, 20, 24 |
| 0.25 | 32, 4, 25, 26, 33, 24, 1, 16, 9, 28, 19, 18, 10, 8, 17, 23, 6, 12, 30, 13, 11, 5, 14, 7, 15, 31, 20, 3, 2, 29, 22, 27, 21 |
| 0.5 | 32, 1, 28, 3, 5, 6, 13, 10, 4, 19, 30, 2, 8, 25, 7, 12, 24, 31, 23, 33, 17, 18, 16, 9, 22, 26, 11, 15, 21, 20, 27, 14, 29 |
| 0.75 | 32, 28, 6, 31, 3, 30, 24, 7, 25, 19, 16, 8, 4, 18, 17, 33, 1, 26, 20, 2, 22, 23, 15, 10, 13, 12, 27, 9, 11, 14, 29, 21, 5 |
| 1.0 | 32, 25, 24, 6, 16, 4, 28, 17, 7, 3, 30, 33, 2, 19, 1, 13, 18, 31, 22, 8, 23, 11, 9, 10, 27, 26, 15, 29, 12, 21, 20, 5, 14 |

由表 2 可知，不同权重因子所对应的排序后的序列特征子集的次序不同。这表明，权重因子能够细致地区分特征冗余性对结果影响的程度，且通过调整权重因子可以改变特征的"价值"高低次序。

然后，采用 SVM 分类器测试特征子集，并按照特征排序结果逐一添加特征，所得分类正确率的变化情况如图 3 所示。

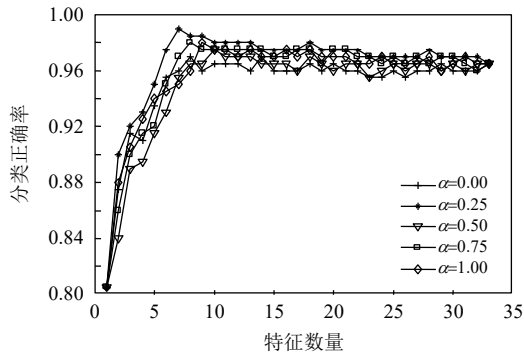

**图 3 不同权重因子对应的分类正确率曲线**
**Fig. 3 Classification accuracy curves corresponding to different weighting factors**

从图 3 可以看到：随着特征数量的增加，分类正确率首先逐渐增大，当达到一个峰值后，将基本保持不变或下降。这表明峰值之后所又增加的特征并没有改善评估模型的分类性能，有些反而给分类带来了不利影响。这一结果表明在模式识别问题中特征选择的必要性，同时也说明文献[17]中指出的特征选择过拟合现象在暂态稳定评估特征选择问题中是同样存在的。

最后，比较各权重因子对应的候选特征子集组的分类正确率，选择得到最大分类正确率的特征集合，如表 3 所示。

**表 3 不同权重因子对应的最大分类正确率特征集合**
**Tab. 3 Feature sets of maximum classification accuracy corresponding to different weighting factors**

| 权重因子 | 分类正确率/% | 特征子集 |
| --- | --- | --- |
| 0 | 97.00 | 32, 10, 4, 19, 1, 9, 28, 16 |
| 0.25 | 99.00 | 32, 4, 25, 26, 33, 24, 1 |
| 0.5 | 97.50 | 32, 1, 28, 3, 5, 6, 13, 10, 4, 19 |
| 0.75 | 98.00 | 32, 28, 6, 31, 3, 30, 24, 7 |
| 1 | 98.00 | 32, 25, 24, 6, 16, 4, 28, 17, 7 |

由表 3 可知，所提方法的最大分类正确率在权重因子为 0.25 时取得，因此，所求的最优特征子集：$A_1$={Tz32,Tz4,Tz25,Tz26,Tz33,Tz24,Tz1}。进一步比较权重因子为 0.5 和 0.25 的对应结果可知：标准 mRMR 在引入权重因子后，不仅最大分类正确率由原来的 97.50% 提高至 99.00%，而且特征子集的维数也由 10 维降为 7 维。这说明，改进 mRMR 能细致刻画特征相关性和冗余性权重，有效去除无关冗余特征，在降维同时也提高了分类正确率，从而验证了改进 mRMR 通过引入权重因子以细化对特征相关性和冗余性度量的有效性。同时，当权重因子分别取 0.75 和 1 时所对应的分类正确率相等，均为 98.00%，而它们所对应的特征子集却不同，这表明不同的特征组合可以具有相同的分类能力。

为了验证所提方法的有效性，将所得特征子集 $A_1$ 与原始特征集 $A$ 及应用主成分分析法(principal components analysis，PCA)所得特征子集 $A_2$ 进行了对比测试，结果见表 4。采用 PCA 方法时，保留原始数据集 95% 的方差，可将原始特征集降维为 10 维的特征子集 $A_2$。

训练暂态稳定评估模型时，SVM 的核函数选用径向基核函数，算法参数通过网格搜索结合 5-折交叉验证优化选取[6]，其中正则化参数 $C$ 和核参数 $\gamma$ 的搜索范围分别为 $[2^{-5}, 2^{15}]$ 和 $[2^{3}, 2^{-15}]$。训练完毕后，利用训练好的最优评估模型对测试集进行测试。考虑到测试准确率 $a_{test}$ 存在一定的偶然性，测试中需兼顾一些具有统计意义的指标[12]，如 Kappa

**表 4 新英格兰 39 节点测试结果**
**Tab. 4 Test results in New England test system**

| 特征集 | 维数 | 模型参数 $C$ | 模型参数 $\gamma$ | 测试正确率/% | $K$ | $r$ | $\eta$ |
| --- | --- | --- | --- | --- | --- | --- | --- |
| $A$ | 33 | 512.0 | 0.5 | 96.50 | 0.929 | 0.961 5 | 0.951 8 |
| $A_1$ | 7 | 32.0 | 0.125 | 99.00 | 0.980 | 0.991 8 | 0.987 3 |
| $A_2$ | 10 | 128.0 | 0.125 | 96.00 | 0.918 | 0.970 8 | 0.949 6 |

统计值 $K$ 和接受者操作特征曲线下方面积 $r$。其中，$K$ 用于衡量对数据集预测分类和实际分类之间的一致性。$r$ 为评价分类器整体性能的常用指标(如果分类模型完美，则 $r=1$；如果模型仅是随机猜测模型，则 $r=0.5$；模型越优，则 $r$ 越大)。为兼顾上述 3 种分类性能指标，本文参照文献[12]，采用 TSA 测试分类综合指标 $\eta$ 以全面客观地评价评估模型的分类性能，其计算公式为

$$\eta = \frac{a_{\text{test}} + K + r}{3} \tag{8}$$

由表 4 可知，采用本文所提方法所得特征子集 $A_1$ 与原始特征集 $A$ 相比具有更优的分类性能($\eta$ 更大)，但维数压缩到原来的 1/4；而且 $A_1$ 比采用 PCA 法所得特征子集 $A_2$ 的分类性能更好，且维数更少。

### 4.2 IEEE 50 机测试系统

#### 4.2.1 样本集的构造

IEEE 50 机测试系统由 50 台发电机、145 条母线和 453 条传输线构成[12,18]。发电机 1—5 采用 4 阶模型，且励磁均采用 IEEE DC1 型励磁系统模型，发电机 6—9 采用 6 阶模型且配置了简单励磁模型，其余各发电机采用经典模型，负荷模型为恒阻抗模型；故障类型为三相短路，故障清除时间为 0.1 s，故障清除后系统拓扑结构不变；系统在从 80%以 5%递增到 105%共计 6 个负荷水平下，每个负荷水平下随机设置 5 种发电机出力，选择 20 个不同的故障位置；所用暂稳判据及仿真软件均与新英格兰 39 节点测试系统相同，共生成 600 个样本，随机选取其中 400 个构成训练集，其余为测试集。

#### 4.2.2 结果与讨论

首先，使用所提方法求取不同权重因子对应的序列特征子集并按 $\Phi(D,R)$ 降序排列，如表 5 所示。

然后，使用 SVM 分类器验证各候选特征子集

表 5  特征降序排列分布
Tab. 5  Feature distribution in descending order

| 权重因子 | 特征排序 |
| --- | --- |
| 0 | 32, 10, 4, 19, 24, 9, 28, 2, 33, 26, 16, 18, 15, 21, 17, 23, 12, 31, 1, 8, 6, 13, 5, 3, 14, 7, 22, 27, 30, 25, 11, 20, 29 |
| 0.25 | 32, 26, 19, 33, 9, 4, 1, 24, 31, 18, 16, 25, 17, 10, 7, 30, 28, 2, 23, 8, 13, 11, 5, 6, 21, 15, 14, 3, 12, 20, 22, 29, 27 |
| 0.5 | 32, 1, 31, 3, 5, 33, 9, 10, 18, 19, 4, 25, 26, 16, 7, 12, 24, 28, 23, 6, 17, 2, 30, 13, 22, 8, 11, 15, 21, 29, 27, 14, 20 |
| 0.75 | 32, 28, 6, 31, 9, 33, 24, 1, 25, 16, 4, 17, 19, 18, 11, 7, 10, 26, 22, 2, 29, 8, 15, 5, 13, 12, 30, 23, 3, 14, 20, 21, 27 |
| 1.0 | 32, 25, 24, 26, 19, 4, 31, 17, 1, 16, 28, 33, 9, 7, 13, 18, 10, 22, 8, 23, 11, 2, 6, 30, 29, 15, 27, 12, 21, 14, 20, 5, 3 |

的分类性能，得到不同权重因子对应最大分类正确率的特征集合，如表 6 所示。

表 6  不同权重因子对应的最大分类正确率特征集合
Tab. 6  Feature sets of maximum classification accuracy corresponding to different weighting factors

| 权重因子 | 分类正确率/% | 特征子集 |
| --- | --- | --- |
| 0 | 95.00 | 32, 10, 4, 19, 24, 9, 28, 2, 33, 26 |
| 0.25 | 96.50 | 32, 26, 19, 33, 9, 4, 1, 24, 31, 18, 16 |
| 0.5 | 95.50 | 32, 1, 31, 3, 5, 33, 9, 10, 18 |
| 0.75 | 96.00 | 32, 28, 6, 31, 9, 33, 24, 1, 25, 16 |
| 1 | 95.50 | 32, 25, 24, 26, 19, 4, 31, 17, 1, 16, 28, 33 |

由表 6 可知，算例 2 中最大分类正确率也在权重因子为 0.25 时取得，此时对应的特征子集为：$B_1=\{Tz32,Tz26,Tz19,Tz33,Tz9,Tz4,Tz1,Tz24,Tz31,Tz18,Tz16\}$。

为了验证所得特征子集的性能，将其与原始特征集 $B$、PCA 法所得降维特征集 $B_2$(保留 95%的方差)及算例 1 中所得推荐特征子集 $B_3=\{Tz32,Tz4,Tz25,Tz26,Tz33,Tz24,Tz1\}$ 进行了对比测试，结果见表 7。

表 7  IEEE 50 机测试系统的测试结果
Tab. 7  Test results in IEEE 50-generator test system

| 特征集 | 维数 | 模型参数 | | 测试正确率/% | $K$ | $r$ | $\eta$ |
| --- | --- | --- | --- | --- | --- | --- | --- |
| | | $C$ | $\gamma$ | | | | |
| $B$ | 33 | 2048.0 | 2.0 | 95.00 | 0.897 | 0.9466 | 0.9312 |
| $B_1$ | 11 | 32.0 | 0.125 | 96.50 | 0.929 | 0.9619 | 0.9520 |
| $B_2$ | 15 | 8192.0 | 0.5 | 94.50 | 0.889 | 0.9499 | 0.9280 |
| $B_3$ | 7 | 512.0 | 0.03125 | 95.00 | 0.898 | 0.9481 | 0.9320 |

由表 7 可知，本文方法所得特征子集 $B_1$ 比原始特征集 $B$ 和采用 PCA 法所得特征子集 $B_2$ 分类性能更优($\eta$ 更大)，且维数减少为 $B$ 的 1/3。算例 1 所得特征子集 $B_3$ 比 $B_1$ 的评估平均指标 $\eta$ 下降了 2.0%，这一结果与文献[12]中的结果基本一致。这说明随着系统规模增大，系统的复杂性随之增大，特征量也应相应地增加，以便为描述系统的稳定特性提供更加完备的材料。

### 4.3 其他暂态稳定评估模型的评估结果

为验证所提特征选择方法的普适性，本文采用多层感知器(multilayer perception，MLP)[2]、概率神经网络(probabilistic neural network，PNN)[4]和极限学习机(extreme learning machine，ELM)[19]等评估模型在得到的特征子集上进行了对比测试，测试结果如表 8 所示。其中，MLP 为单隐层网络，神经元个数为 15，训练算法为反向传播算法，学习率 0.8，

动量因子为0.7；PNN 的径向基函数传播系数为0.1；ELM 的隐层节点个数为50。

由表 8 可知，MLP、PNN 和 ELM 等 3 种评估模型在本文所提方法所得特征子集的分类性能均与原始特征集相似，表明所提方法同样适用于其它暂态稳定评估模型。

表 8 其它模型的测试结果
Tab. 8 Test results of other models

| 测试算例 | 评估模型 | 特征集 | 测试正确率/% | $K$ | $r$ | $\eta$ |
|---|---|---|---|---|---|---|
| 新英格兰39节点系统 | MLP | A | 96.00 | 0.919 | 0.983 1 | 0.954 0 |
| | | $A_1$ | 95.00 | 0.899 | 0.972 2 | 0.940 4 |
| | PNN | A | 97.00 | 0.939 | 0.984 7 | 0.964 6 |
| | | $A_1$ | 96.50 | 0.929 | 0.975 1 | 0.956 4 |
| | ELM | A | 98.50 | 0.969 | 0.993 1 | 0.982 4 |
| | | $A_1$ | 98.00 | 0.959 | 0.983 0 | 0.974 0 |
| IEEE 50机测试系统 | MLP | B | 92.00 | 0.839 | 0.932 0 | 0.897 0 |
| | | $B_1$ | 91.50 | 0.829 | 0.924 1 | 0.889 4 |
| | PNN | B | 95.50 | 0.908 | 0.959 6 | 0.940 9 |
| | | $B_1$ | 95.00 | 0.898 | 0.950 0 | 0.932 7 |
| | ELM | B | 97.00 | 0.939 | 0.963 5 | 0.957 5 |
| | | $B_1$ | 96.50 | 0.928 | 0.952 6 | 0.948 5 |

## 5 结论

针对电力系统暂态稳定评估的特征选择问题，本文提出了一种基于改进 mRMR 和支持向量机的特征选择方法，并在新英格兰 39 节点系统和 IEEE 50 机测试系统上进行了仿真研究，结论如下：

1）改进 mRMR 通过引入特征相关冗余权重因子能有效细化对特征相关性和冗余性的平衡和度量，且其性能比标准 mRMR 更优；

2）在保持分类性能不降低情况下，所提方法能大幅降低原始样本集的维数，且优于常用的主成分分析法；

3）所提出的特征选择方法同样适用于其他暂态稳定评估模型，如多层感知器、概率神经网络和极限学习机等。

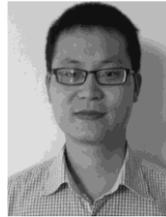
李扬

收稿日期：
作者简介：
李扬(1980)，男，博士研究生，电力系统安全稳定评估与控制、智能技术在电力系统中的应用；

顾雪平(1964)，男，博士，教授，博士生导师，主要研究方向为电力系统安全防御和恢复控制、电力系统安全稳定评估与控制、智能技术在电力系统中的应用。